# Detection of a Red Supergiant Progenitor Star of a Type II-Plateau Supernova

Stephen J. Smartt†, Justyn R. Maund†, Margaret A. Hendry†, Christopher A. Tout†, Gerard F. Gilmore†, Seppo Mattila*, Chris R. Benn¶

*†Institute of Astronomy, University of Cambridge, Madingley Road, Cambridge CB3 0HA, UK.*

*\*Stockholm Observatory, AlbaNova University Center, SE-106 91, Stockholm, Sweden*

*¶ Isaac Newton Group of Telescopes, Apartado 321, Santa Cruz de La Palma, E-38700 Spain*

**We present the discovery of a red supergiant star that exploded as supernova 2003gd in the nearby spiral galaxy M74. The Hubble Space Telescope (HST) and the Gemini Telescope imaged this galaxy 6 to 9 months before the supernova explosion and subsequent HST images confirm the positional coincidence of the supernova with a single, resolved star which is an $8^{+4}_{-2}$ solar mass red supergiant. This confirms both stellar evolution models and supernova theories which predict that type II-Plateau supernovae have cool red supergiants as their immediate progenitor stars.**

Supernova 2003gd was discovered on June 12.82 UT in the nearby spiral galaxy M74 *(1)*. It was rapidly shown to be a type II-Plateau (II-P) supernova which was discovered about 87 days after explosion *(2-5)*. The progenitors of type II-P supernovae have long been thought to be red supergiant stars with initial masses greater than 8-10M$_\odot$ that have retained their hydrogen envelopes before core-collapse. This model accounts for the 2 to 3 month long plateau phases seen in the lightcurves of SNe II-P, the existence of hydrogen P-Cygni profiles in the early time spectra and the estimated physical parameters of the expanding photosphere such as velocity, temperature and density (*6-8*). Stellar evolutionary calculations are consistent with this picture where stars with initial masses in the range 8-25$M_\odot$ reach the end of their nuclear burning lives when they are red supergiants (*9-11*). Only two progenitors of unambiguous supernovae have been directly identified and yielded estimates of luminosity, temperature and mass (see *12* for a review). These are the progenitors of the peculiar type II-P supernova 1987A which was a blue supergiant (13, 14), and the IIb supernova 1993J that arose in a massive interacting binary system (15, 16). The expected red supergiant origin for the common type II-P supernovae has so far eluded direct detection. Furthermore, recent extensive analyses of the physical parameters of these supernovae have inferred progenitor masses and radii for some events which are inconsistent with the progenitor being an M-type supergiant (8). The fortuitous coincidence of a II-P supernova occurring in a nearby galaxy which has high quality prediscovery images available has allowed us to directly determine the physical parameters of a supernova progenitor for only the third time.

The galaxy M74 was observed with the Wide Field and Planetary Camera 2 (WFPC2) on board the Hubble Space Telescope (HST) through the F606W and F300W filters about 200 days before the estimated explosion date of supernova (SN) 2003gd of 18 March 2003 (which has an uncertainty of approximately 21 days). This galaxy was also observed about 310 days pre-explosion by the Gemini Telescope during system verification of the Gemini Multiobject Spectrograph (GMOS) in the Sloan Digital Sky Survey (SDSS) like filters g', r' and i'. On 1 and 2 August 2003 we observed the supernova with the Hubble Space Telescope, this time using the High Resolution Channel (HRC) of the Advanced Camera for Surveys (ACS) (Table1). An approximate position for the SN was estimated from ground based images (*4, 17*), however the image resolution and accuracy of the astrometry were not good enough to definitively identify a progenitor star. We have performed precise differential astrometry using the ACS

F555W image as the reference frame. Sixteen bright point sources were simultaneously identified in the ACS image and the WFPC2 images (in a 26 arcsecond field centered on the SN) and a geometric spatial transformation function was computed to map the WFPC2 coordinate system onto the ACS reference (*12*). Ten point sources were unambiguously resolved and identifiable in both the GMOS and ACS images. We identify an object in the pre-explosion WFPC2 F606W and GMOS *i*' images that is coincident with SN 2003gd (Fig. 1). There is no object visible in the F300W (limiting magnitude $m_{300}$=23.9) or the GMOS *g*' image (limiting magnitude *g*' =25.3) but there is a faint object visible in the GMOS *r*' which is coincident with the SN and is close to the 3σ detection limit of the image. The positional difference between SN 2003gd and the progenitor object identified in the WFPC2 F606W filter is 13±33 milliarcseconds (Table 2). Hence within the error estimates of the differential astrometric solution the progenitor object identified in the WFPC2 image is coincident with the position of SN 2003gd. This star is the object initially discovered in (*17*) and is Star A in (*4*). We stress that it is virtually certain that we have identified the progenitor. There is certainly no other possible object in the WFPC2 frames and if the progenitor is undetected it would have a magnitude fainter than $V\approx27.1$ (the 3σ limit) or $M_V$=–2.7 (at a distance of 9Mpc). This would correspond to around 5M$_\odot$ which is too low for core-collapse to occur. The resolution of the GMOS *i*' image (0.57 arcseconds) is lower than that of the space based observations, and the difference in position of the progenitor source and SN is 137±71 milliarcseconds. Within the uncertainties this is not exactly coincident with the SN suggesting that the GMOS *i*' source is a blend of the Stars A, C and D (Fig. 1). We have utilised the Image Reduction and Analysis Facility (IRAF) implemtation of the CPLUCY image restoration algorithm (18) to separate the constituents of the blended GMOS *i'* source. The positions of stars A, C and D on the WFPC2 frame were transformed to the GMOS *i'* frame and these positions were used as the coordinates of the individual flux sources. The deconvolution process yielded a A:C flux ratio of 4±2:1. The flux contribution from Star D was negligible.

The magnitude of the progenitor in the WFPC2 image was determined using the photometry package HSTphot (*19,20*).We initially measured a magnitude in the F606W filter and converted this to the standard Johnson *V* filter by assuming an intrinsic *V–I* color and applying the standard WFPC2 transformations (*21*) The I-band magnitude was calculated as detailed below and this iterative procedure produced a value of *V*=25.8±0.15 (supporting online text). Magnitudes of stars in the GMOS *i'* filter were calculated using the zero-points provided in the release of this calibrated dataset, and were tied to the Johnson I-band system (*I*) images from the Isaac Newton Telescope (*22*) and the standard sequence used to monitor the supernova 2002ap (*23*). The total magnitude for the object is *I*=23.13±0.13, where the error is the combination of the color transformation error (0.11) in the method, and the flux measurement error (0.07) of the object. In a similar manner we measure the faint source on the *r'* band image to have a standard Johnson magnitude *R*=25.0±0.5. From the astrometric considerations it is likely that the GMOS sources are constructed of the progenitor and Star-1 in a ratio of 4:1, hence the *RI* magnitudes should be increased by 0.2±0.1.

The distance to the galaxy has previously been determined to be 7.5±0.5 Mpc from two studies of the brightest stars in M74 and its satellites (*24,25*). However this method is prone to uncertainties, particularly as ground based imaging was used which may mistake compact clusters for individual massive stars, and hence systematically underestimate the distance. For example the distance estimate to another galaxy (NGC1637) using this method may be seriously underestimated as shown by a new distance determination using Cepheid variable stars (*26*). We have used the detailed photometric and spectroscopic monitoring of SN 2003gd (supporting online text) to calculate a distance to M74 using the Standard Candle Method (*27,28*) and combine this

5with four other determinations of distance in the literature to get a mean value of 9.1±1.9 Mpc (Table 3). We estimate the reddening to the supernova from three distinct methods which are color evolution of the supernova, the three color photometry of stars close to the supernova in the ACS images (supporting online text), and the extinction from the emission lines of neighboring HII regions (Table 3). We estimate a reddening of $E(B-V)$=0.14±0.13, and if we assume a standard Galactic extinction law with $R$=3.1 (*29*), this corresponds to a total visual extinction of $A_v$=0.43±0.40. The absolute bolometric magnitude ($M_{bol}$) of the progenitor is then calculated with the equation

$$M_{bol} = 5 - 5\log d - A_v + V + BC,$$

where $d$ is the distance in parsecs, $V$ is the visual magnitude and BC is the bolometric correction. From the $V$ and $I$ magnitudes discussed above, the apparent color of the progenitor star is $(V-I)$=2.5±0.2 and so the standard reddening law (*29*) implies $(V-I)_0$=2.3±0.3. The intrinsic colors of K and M-type supergiants (*30*) imply that the progenitor star has a spectral type in the range K5-M3Ib. Although the progenitor's $R$-band magnitude has a large error, the intrinsic color $(R-I)_0$=1.8±0.5 is consistent with this spectral type. Applying an appropriate bolometric correction BC=–1.5±0.5 (*30,31*) results in $M_{bol}$ = –5.9±0.8, and assuming a solar bolometric magnitude of +4.74, this corresponds to a luminosity for star A of $\log L/L_\odot$ =4.3±0.3 (note that the errors on all of the factors on the right hand side of the equation are combined (in quadrature) to give this error on luminosity) The temperature scale of early-M supergiants would suggest the progenitor has an effective temperature of $T_{eff}\approx$3500K (*31*), which allows the progenitor to be placed in the Hertzprung-Russell diagram to compare with stellar evolutionary tracks (Fig. 2). The progenitor star's colors and luminosity are consistent with it being a red supergiant which had an initial mass on the main sequence of $8^{+4}_{-2}M_\odot$. This is at the lower end of the mass range in which we expect stars to develop a Chandrasekhar mass core and be capable of undergoing core-collapse.

This is the first detection of a progenitor star from a normal type II-P supernova, which is the most common type of supernova (by volume) in the Universe. It is a red supergiant which is consistent with the models of single stellar evolution. Recently there have been attempts to identify the progenitors of three nearby type II-P supernovae by the same methods as described here. Although these have failed to detect an object, they have been able to set restrictive upper mass limits. Mass limits of the progenitors were estimated to be $\leq 15 M_\odot$ for SNe 1999em and 2001du and $\leq 12 M_\odot$ for SN 1999gi (*11,12,32*). These three supernovae and 2003gd are all spectroscopically very similar and appear to be a common, homogeneous class of type II. Stellar evolutionary models and theories of the supernovae lightcurve and spectral evolution have long predicted that red supergiants should be the progenitors of SNe type II-P. We have directly confirmed these stellar evolutionary models for SN 2003gd and the upper mass limits set for three other very similar supernovae are consistent with this scenario. However there is a quantitative discrepancy now appearing between the masses derived for these four SNe II-P and the mass required to support the long plateau phase with normal expansion velocities. Consistently high ejecta masses have been derived for a large sample of 13 SNe II-P in the range $17-56 M_\odot$ (*8*), quite different from the low masses that the direct method suggests. The three SNe with excellent monitoring data and direct mass limits do show agreement however (*12*) which is strong evidence that the common type II-P supernovae originate in stars with masses between $8-15 M_\odot$. This is at the lower end of the range of masses for which core collapse is expected to occur and this is in agreement with the latest pre-supernova stellar models (*9*).

77

**TABLE 1**: The details of all the observational data. Estimated supernova explosion date is 17 March 2003 (supporting online text).

| Telescope and instrument | Dataset | Date | Filter | Exposure Time |
|---|---|---|---|---|
| HST/WFPC2 | U81XCA01B | 25 August 2002 | F606W | 1000s |
| HST/WFPC2 | U81XCY01B | 28 August 2002 | F606W | 2100s |
| HST/WFPC2 | U81XCA03B | 25 August 2002 | F300W | 3000s |
|  |  |  |  |  |
| Gemini/GMOS | GN-2001B-SV-102 | 11 May 2002 | $g'$ | 960s |
| Gemini/GMOS | GN-2001B-SV-102 | 11 May 2002 | $r'$ | 480s |
| Gemini/GMOS | GN-2001B-SV-102 | 11 May 2002 | $i'$ | 480s |
|  |  |  |  |  |
| HST/ACS | J8NV01011/21 | 1 August 2003 | F439W | 2500s |
| HST/ACS | J8NV01031/41 | 2 August 2003 | F555W | 1100s |
| HST/ACS | J8NV01051 | 2 August 2003 | F814W | 1100s |

**TABLE 2: Astrometry of post and pre-explosion images.** This table lists the sources of error in the astrometry of the supernova and the progenitor star. The error in the position of the progenitor star in the under-sampled WFPC2 image was estimated from the standard deviation of four methods used to carry out photometry and astrometry. These were the centroid, Gaussian fit, optimal filtering methods (all within the IRAF DAOPHOT package) and point-spread-function fitting (within the HSTphot package – see references 19 and 20). The error in the position of the supernova was determined in a similar way. The geometric transformation error is a combinations of the RMS residuals from the 2-dimension spatial transformation functions. The total error in the differential astrometric solution is calculated by combining these three independent errors in quadrature. A similar procedure was applied to the GMOS $i'$ frames.

|  | **WFPC2** (milliarcseconds) | **GMOS $i'$** (milliarcseconds) |
|---|---|---|
| Position of progenitor error | 31 | 58 |
| Position of SN error | 3 | 3 |
| Geometric transformation (RMS) | 11 | 41 |
| **Total error** | 33 | 71 |
| **Position difference** | 13 | 137 |

**TABLE 3: Compilation of distance estimates of M74 and reddening towards the line of sight of SN2003gd.** We list the individual measurements which are combined to give the best estimate of distance and reddening. The kinematic distance to M74 comes from the heliocentric radial velocity of 656 kms$^{-1}$ and application of three models for the infall onto the Virgo cluster which are all corrected for $H_0$=72 kms$^{-1}$ (*37*). The three values are are 11.1 Mpc (*34*), 10.1 (*38*), 9.5 (*39, 40*). The simple mean is listed below and the uncertainty is the standard deviation of the measurements plus the uncertainty due to the cosmic thermal velocity dispersion of 187 kms$^{-1}$ (*41*). The use of a kinematic distance alone is limited by the magnitude of velocity dispersions around the Hubble Flow seen in galaxies in the local Universe and the application of a accurate model to account for Virgo and the Great Attractor infalls (*41*), hence there is a large uncertainty in this value. The distance from the brightest stars is a combination of two results 7.2±1.7 (*25*) and 7.8±0.4 (*24*). Although the errors quoted are quite small this method suffers from the difficulties of resolving the stars from compact clusters and HII regions (the studies were ground-based), confusion with foreground stars and the few objects at the brightest tip. Indeed an extensive study of the use of the method suggests a realistic error in deriving distance modulii is 0.55$^m$ (*42*). We use the mean of the two distance estimates with an error determined from the statistical uncertainty of 0.6 Mpc and systematic uncertainty of 2.2 Mpc. Our new distance from the Standardized Candle Method for type II-P supernovae results in a distance of 9.7±3.5 Mpc. There are drawbacks with all three of these methods and all have similar uncertainties. It would seem appropriate to combine the three to give an unweighted mean in order to derive the luminosity of the progenitor star of 2003gd. The extinction toward the line of sight of the supernova and the progenitor can be estimated in three ways. The recent study of the SN2003gd lightcurve (*4*) produced a reddening estimate toward the supernova of $E(B-V)$=0.13±0.03. We have used an identical method to determine the reddening from the data in Fig. S1 (supporting online material), and derive the same result but with a larger error $E(B-V)$=0.11±0.16. The extinction measured from the colors of surrounding stars in the HST ACS multi-color images (see Fig. S3 in supporting online material) also gives a consistent result. Similar to the distance calculation, all three of these methods have drawbacks and advantages. The supernovae 1999em and 2003gd may show intrinsically different color evolution, the HII regions may show different extinction estimates than their ionising stars (*12* and references therein), there may be large local reddening variation toward the nearby stars. However these estimates are the best that can be achieved with the present data and it is encouraging that they are very similar. A mean of the three methods gives $E(B-V)$=0.14±0.13. The uncertainty comes

not from the standard deviation of the sample (at 0.04 this is unrealistically small), but from a mean of the uncertainties in the three methods.

| Distance (Mpc) | Method (reference) | E(B−V) | Method (reference) |
|---:|---|---:|---|
| 7.5 ± 2.8 | Brightest supergiants | 0.11 ± 0.16 | SN color evolution |
| 10.2 ± 3.4 | Kinematic | 0.19 ± 0.15 | Nearby HII regions |
| 9.7 ± 3.5 | Type II SN Standard Candle Method | 0.13 ± 0.07 | Color of nearby stars |
| **Mean =9.1±1.9** | | **Mean=0.14 ± 0.13** | |

**Figure 1 (Caption):** The pre and post-explosion images from HST ACS, HST WFPC2, and Gemini GMOS instruments (for details see Table 1). **(A)** The supernova 2003gd is marked in the ACS image which has a spatial resolution of 0.05 arcseconds. **(B)** The middle panel shows the progenitor (Star A) identified in the WFPC2 F606W filter image, this is the total co-added 3100s image and the object is detected at 10$\sigma$ above the background noise. It has the profile of an unresolved instrument point-spread-function, and is hence a single object at the resolution of the WFPC2 camera (around 0.15 arcseconds). The positions of the progenitor star and supernova coincide within 13±33 milliarcseconds (see Table 2 for discussion of errors). This is the same star A as in reference (*4*) and we have kept the same nomenclature for clarity. We note the position of their star B and label the additional nearby objects C and D. **(C)** The ground-based image from GMOS in the *i'* band which has a spatial resolution of 0.57 arcseconds. When differential astrometry is applied to the GMOS frame to transform it to the ACS coordinate grid, the difference in the position of the progenitor and the *i'*-band source is 137±71 milliarcseconds. We can model this source with two singular components at the measured WFPC2 positions of the stars A and C with a flux ratio of 4±2:1.

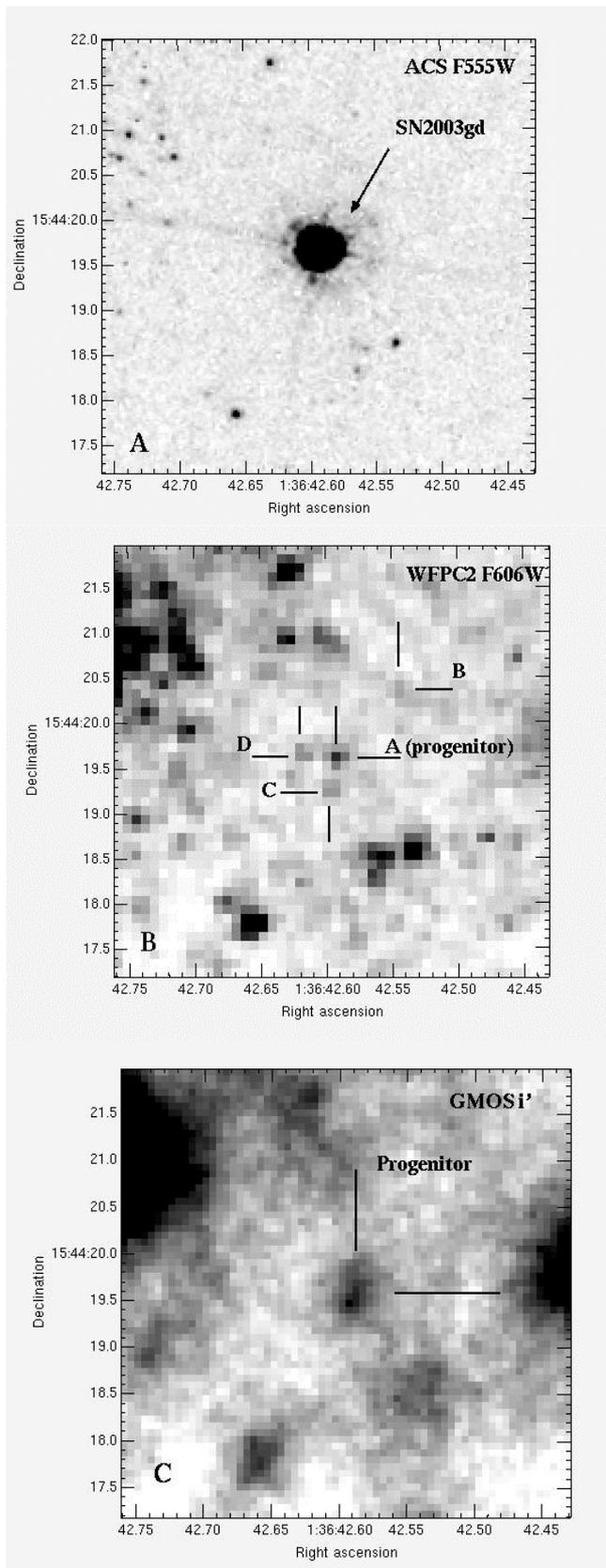

Fig. 1

**Figure 2 (Caption):** The position of the stellar progenitor of SN 2003gd is shown on a theoretical HR-diagram with the stellar evolutionary tracks shown. These tracks show the evolution of stars of masses 8-20$M_\odot$ and are marked with the initial stellar mass accordingly (for full details of the models see *12,33*). A solar metallicity was chosen for the tracks as implied by recent analysis of the abundance gradient from HII regions in the galaxy M74 and the galactocentric radius of the supernova (*34*). The progenitor star's position is close to the red supergiant branch of a star that had an initial mass of 8$M_\odot$. However the error ellipse does overlap a substantial part of the red supergiant branch of a 10$M_\odot$ star during which the star will undergo carbon burning. The mass-luminosity relation of red-supergiants is somewhat uncertain but the position of the progenitor is consistent with a red supergiant of 8-10$M_\odot$. In Fig. S4 (in Supporting Online Material) we compare results of two other contemporary model grids and found that the end points of our tracks for stars in the 8-15$M_\odot$ lie at the same temperatures and within ±0.2 dex in luminosity (*35,36*). At a fixed luminosity the mass that one derives could differ by a maximum 2$M_\odot$. Hence this suggests we should conservatively increase the mass range inferred by comparison with the models to 8-12$M_\odot$. The lowest mass star that can support a successful supernova is around 6-12$M_\odot$ (supporting online text). Hence if we take the lower bound conservatively to be 6$M_\odot$ then the mass range allowed for the progenitor is $8^{+4}_{-2}M_\odot$. At these masses the mass-loss rate is low and inclusion does not seriously affect the pre-supernova positions. We also compared the pre-supernova evolution of rotating stellar models from the same groups and in this mass range rotation rates of up to about 300 kms$^{-1}$ did not affect the point of core collapse by more than ±0.1 dex in luminosity. Rotation does push the higher mass tracks

(20-25 $M_\odot$) to significantly higher luminosities in the red supergiant stage (by up to 0.3 dex), but these are much too high to be consistent with the absolute magnitude of the observed progenitor.

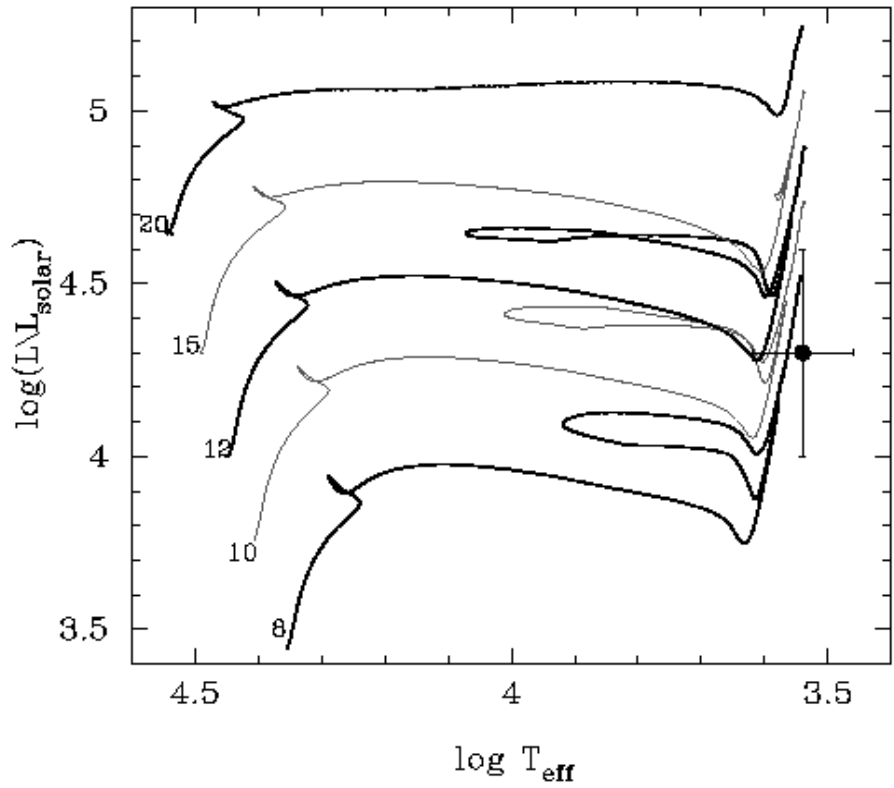

with program GO9733. Also based on observations obtained with the Gemini Telescope. We acknowledge the support given by ASTROVIRTEL, a project funded by the European Commission under FP5 Contract No. HPRI-CT-1999-00081. JRM, MAH and SJS thank PPARC for financial support and SM acknowledges the 'Physics of Type Ia SNe' RTN under contract HPRN-CT-2002-00303. We thank Bill Januszewski for help in coordinating and executing the HST observations and Peter Meikle and Rubina Kotak for assistance, advice and reading the manuscript.

**Correspondence and requests for materials should be addressed to sjs (e-mail: sjs@ast.cam.ac.uk).**